\documentclass{aa}
\usepackage{graphicx}
\usepackage{txfonts}
\newcommand{\eg}{{\it e.g.}, }
\newcommand{\ie}{{\it i.e.}, }
\newcommand{\ms}{m\,s$^{\rm -1}$}
\newcommand{\kms}{km\,s$^{\rm -1}$}
\newcommand{\Mjup}{M$_{\rm Jup}$}
\newcommand{\Msun}{M$_{\sun}$}
\newcommand{\vsini}{$v\sin{i}$}
\newcommand{\bv}{$B-V$}
\newcommand{\harps}{H{\small ARPS}}

\begin{document}
   \title{Extrasolar planets and brown dwarfs around A--F type stars
     \thanks{Based on observations collected at the European Organisation for Astronomical Research in the Southern Hemisphere, Chile, ESO. Program IDs 072.C-0488, 076.C-0279, 077.C-0295, 078.C-0209, 080.C-0664, 080.C-0712, 081.C-0774.}
   \thanks{Table of radial velocities is only available in electronic format at the CDS via anonymous ftp to cdsarc.u-strasbg.fr (130.79.128.5) or via http://cdsweb.u-strabg.fr/cgi-bin/qcat?J/A+A/}}

   \subtitle{V. A planetary system found with HARPS around the F6IV--V star \object{HD\,60532}.}

   \author{
     M. Desort \inst{1} \and
     A.-M. Lagrange \inst{1} \and
     F. Galland \inst{1} \and
     H. Beust \inst{1} \and
     S. Udry \inst{2} \and
     M. Mayor \inst{2} \and
     G. Lo Curto \inst{3}
   }

   \offprints{
     M. Desort\\
     \email{mdesort@obs.ujf-grenoble.fr}
   }

   \institute{
     Laboratoire d'Astrophysique de Grenoble, UMR5571 CNRS, Universit\'e Joseph Fourier, BP 53, 38041 Grenoble Cedex 9, France
     \and
     Observatoire de Gen\`eve, Universit\'e de Gen\`eve, 51 Chemin des Maillettes, 1290 Sauverny, Switzerland
     \and
     European Southern Observatory, Alonso de Cordova 3107, Vitacura, Santiago, Chile
   }

   \date{Received date / Accepted date}

   
   \abstract
   {}
   {In the frame of the search for extrasolar planets and brown dwarfs around early-type stars, we present the results obtained for the F-type main-sequence star HD\,60532 (F6V) with \harps.}
   {Using 147 spectra obtained with \harps~at {\small La Silla} on a time baseline of two years, we study the radial velocities of this star.}
   {HD\,60532 radial velocities are periodically variable, and the variations have a Keplerian origin. This star is surrounded by a planetary system of two planets with minimum masses of 1 and 2.5\,\Mjup~and orbital separations of 0.76 and 1.58\,AU respectively. We also detect high-frequency, low-amplitude (10\,\ms~peak-to-peak) pulsations. Dynamical studies of the system point toward a possible 3:1 mean-motion resonance which should be confirmed within the next decade.}
   {}
   \keywords{techniques: radial velocities - stars: early-type - stars: planetary systems - stars: oscillations - stars: individual: HD\,60532}

   \maketitle
%

\section{Introduction}

  Radial-velocity (RV) surveys have lead to the detection of nearly 300 planets\footnote{Jean Schneider, http://exoplanet.eu} during the past decade. They mainly focus on solar and late-type main-sequence (MS) stars ($\ga$\,F7) which exhibit numerous lines with low rotational broadening. It is often thought that planets around more massive MS stars are not accessible to radial-velocity techniques, as they present a small number of stellar lines, usually broadened and blended by stellar rotation. However, we recently showed (\cite{galland05a} 2005a, Paper\,I) that with a new radial-velocity measurement method that we developed, it is possible to detect planets even around early A-type MS stars with large rotational velocities (typically 100\,\kms). Finding planets around such massive MS stars is of importance, as this allows to test planetary formation and evolution processes around a larger variety of stars, in terms of stellar mass and time scales of evolution processes. This approach is complementary with the one which intends to detect planets around evolved intermediate-mass stars (\eg \cite{sato05} 2005, \cite{lovis07} 2007). In this case, close-in planets have been wiped out but the stellar variability is in principle less intense.

  We performed a radial-velocity survey dedicated to the search for extrasolar planets and brown dwarfs around a volume-limited sample of A--F main-sequence stars with the \harps~spectrograph (\cite{mayor03} 2003) installed on the 3.6-m ESO telescope at La Silla Observatory (Chile). We monitored a sample of 185 MS stars with \bv~ranging between $-0.1$ and 0.6. From the measured RV jitter, we computed the minimum detectable masses with \harps, and showed that in 100 cases, planets with periods smaller than 100 days can be detected, even around stars with early spectral types (down to $\sim$0.1\,\Mjup~at 100 days around slow-rotating late-F stars). Given the data at hand, we also provided the achieved detection limits on the individual targets (\cite{lagrange08} 2008).

  In the course of this survey, we identified a few stars which RV variations could be attributed to planets. Most of these stars are still under follow-up. We present here the detection of a planetary system around one of these stars, HD\,60532. Section~2 provides the stellar properties of this star, the measurement of the radial velocities and their relevance; we also present a Keplerian solution associated to the presence of two planets.

  \section{Stellar characteristics and measurements}

  \subsection{Stellar properties}

  HD\,60532 (HIP\,36795, HR\,2906) is located at 25.7\,pc from the Sun (\cite{hipparcos97} 1997). Stellar parameters such as mass, age, metallicity, rotational velocity and effective temperature are taken from \cite{nordstrom04} (2004), and the gravity is taken from \cite{gray06} (2006). Those values are reported in Table~\ref{tab:stellar_param}. They are in agreement with a spectral type F6IV--V (F6IV in the Bright Star Catalogue, \cite{hoffleit91} (1991), F6V in the H{\small IPPARCOS} catalogue (\cite{hipparcos97} 1997).


  \begin{table}[t!]
    \caption{HD\,60532 stellar properties. Photometric and astrometric data are extracted from the H{\small IPPARCOS} catalogue (\cite{hipparcos97} 1997); spectroscopic data are from \cite{nordstrom04} (2004).}
    \label{tab:stellar_param}
    \begin{center}
      \begin{tabular}{l l c}
        \hline
	\hline
        Parameter       &         & HD\,60532  \\
	\hline
        Spectral Type   &         & F6IV-V \\
        \vsini          & [\kms]  & 8  \\
        $V$             &         & 4.45 \\
        \bv             &         & 0.52 \\
        $\pi$           & [mas]   & $38.9 \pm 0.7$  \\
        Distance        & [pc]    & 25.7 \\
        $M_V$           &         & 2.40 \\
	$\log R'_{\rm HK}$	&	  & $-$4.94 \\
	$[$Fe/H$]$      &         & $-0.42$ \\
        $T_{\rm eff}$   & [K]     & 6095 \\
        $\log g$        & [cm\,s$^{\rm -2}$] & $-3.83$  \\
        $M_1$           & [\Msun] &  $1.44_{-0.1}^{+0.03}$ \\
	Age             & [Gyr]   &  $2.7 \pm 0.1$  \\
        \hline
      \end{tabular}
    \end{center}
  \end{table}

  \subsection{Radial-velocity measurements}

  Since February 2006, 147 high signal-to-noise ratio ($\mathrm{S/N}$) spectra have been acquired with \harps, with a $\mathrm{S/N}$ equal to 310 on average. Each spectrum is formed by 72 spectral orders covering the spectral window [3800\,\AA, 6900\,\AA], with a resolution $R \approx 115\,000$.

  The radial velocities and associated uncertainties have been measured with a dedicated tool (S{\small AFIR}) which uses the Fourier interspectrum method described in \cite{chelli00} (2000) and in \cite{galland05a} (2005a).
 The uncertainty is equal to 0.9\,\ms~on average, consistent with the value obtained from our simulations (see Paper\,I), and includes photon noise, instrumental effects and guiding errors (fixed at a upper limit of 0.5\,\ms).

Note: given the star spectral type, we could also derive the RV from gaussian adjustements to the CCFs (Cross-Correlation Functions), using a mask with a G2 spectral type. The obtained values are comptatible within the error bars with the ones measured by S{\small AFIR}.

  The amplitude of the radial-velocity variations (more than 120\,\ms) as well as their standard deviation ($\sigma_{\rm RV} = 27$\,\ms) are much larger than the uncertainties. We now show that these variations are not due to line-profile variations and we interpret them in terms of the presence of planetary companions.

\begin{figure}[t!]
  \centering
  \includegraphics[width=1\hsize]{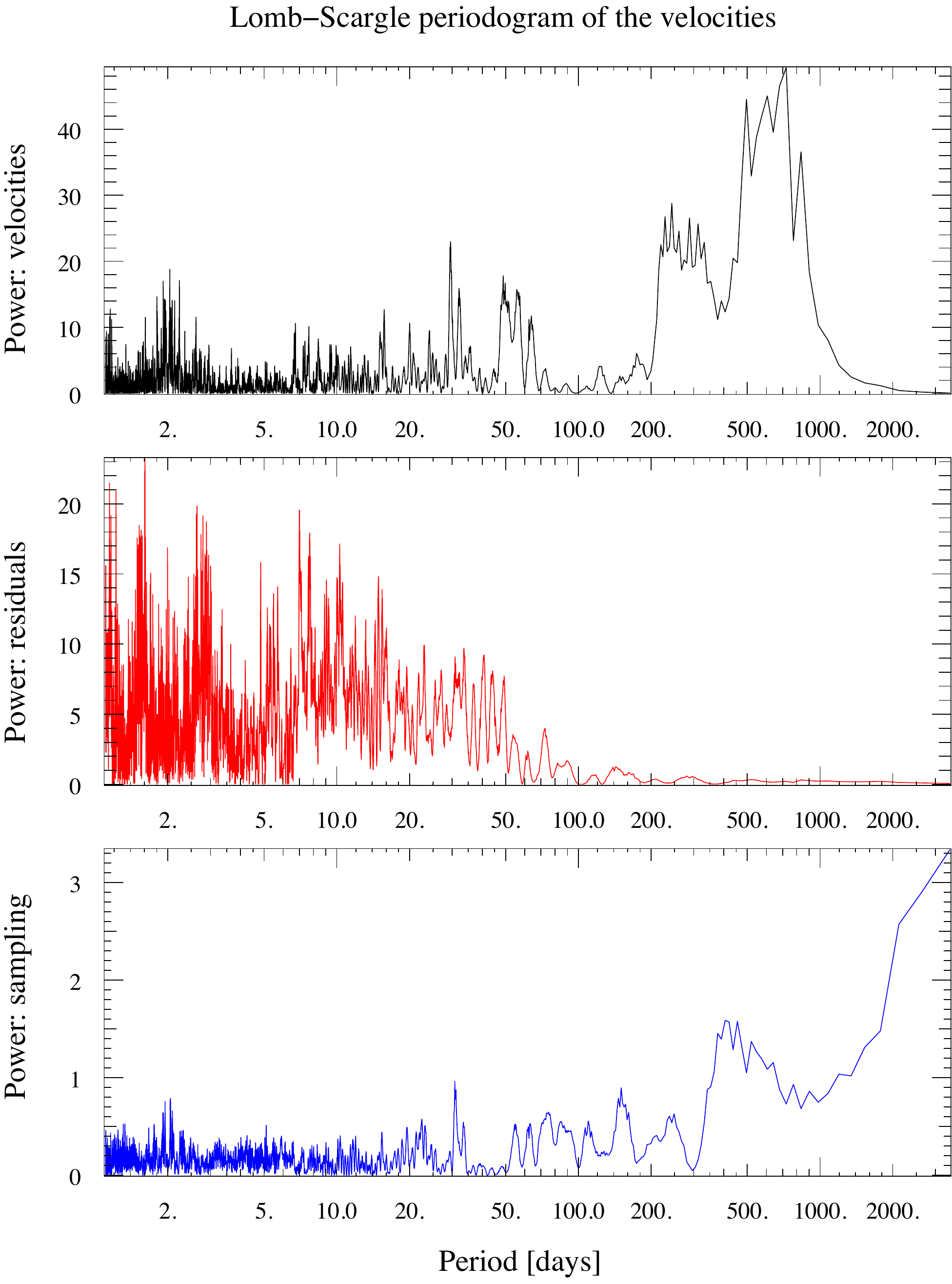}
  \caption{Periodogram of the RVs ({\it top}). Four main peaks are visible. The two near 30 days are observation alias ({\it bottom}, periodogram of the observation epochs), the other two correspond roughly to periods of $\sim$220 and $\sim$500--1000 days. These periods are not well constrained because of the limited phase coverage (our data span only over $\sim$900 days. The middle panel shows the periodogram of the residuals to the final keplerian solution.}
  \label{fig:perio}
\end{figure}

The periodogram (Fig.~\ref{fig:perio}) of the RVs indicates four main periodicities in the data. Two of them (near 30 days) are aliases of data (Fig.~\ref{fig:perio}, {\it bottom} represents the alias generated by our sampling frequency), whereas the other two ($\sim$220 and $\sim$500--1000 days) are real periodicities of the RVs. This points toward the existence of two companions in orbit with periods close to those one.

  \subsection{A Keplerian origin to the main variations}
  \label{sec:kepler_origin}

\begin{figure}[t!]
    \centering
    \includegraphics[width=0.8\hsize]{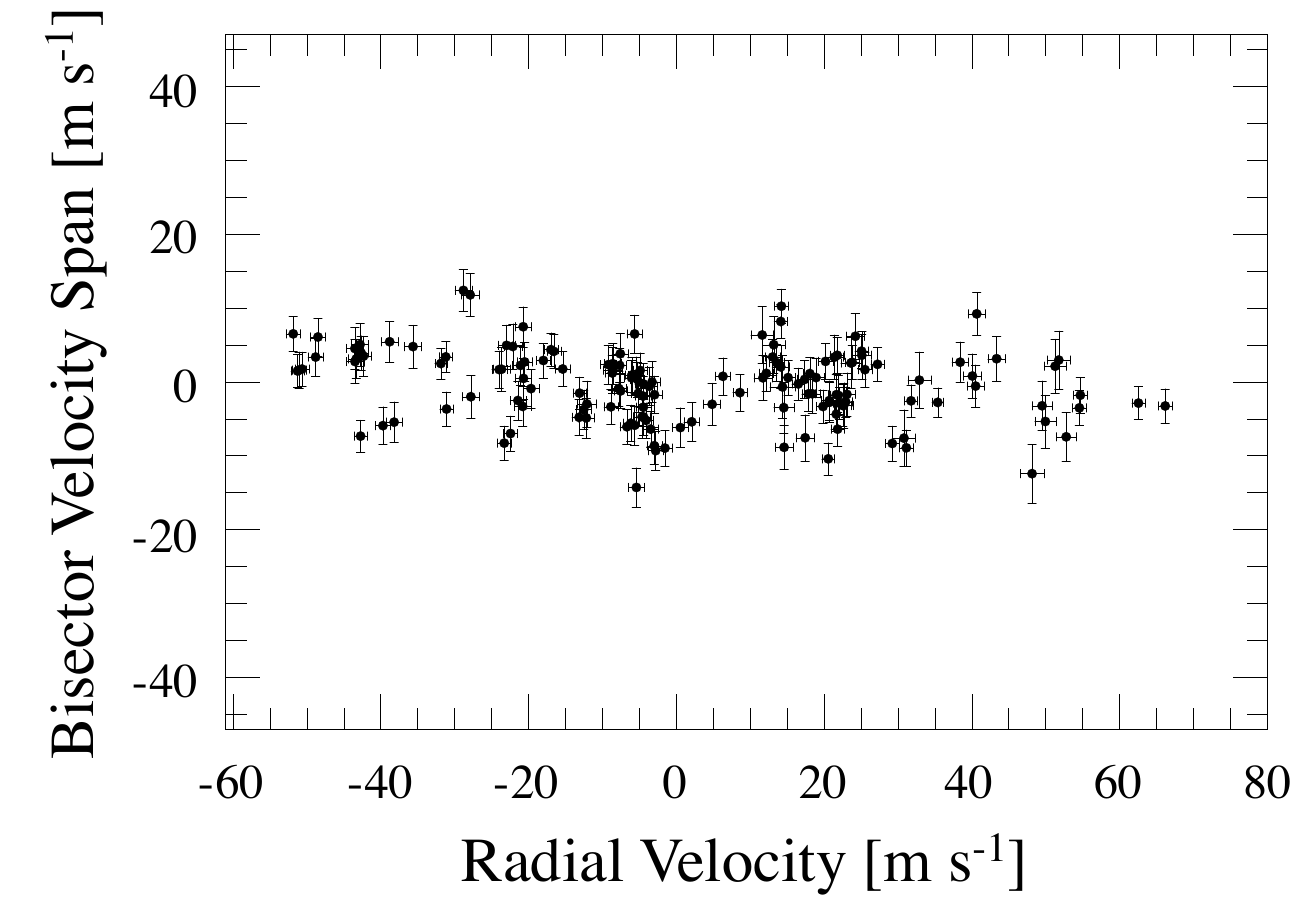}
    \caption{Bisector velocity spans versus RVs. The bisector velocity spans lie within $\pm 15$\,\ms~whereas the RVs vary over more than 120\,\ms~without correlation (see Sect.~\ref{sec:kepler_origin}).}
    \label{fig:bis_span}
  \end{figure}

\begin{figure}[t!]
    \centering
    \includegraphics[width=0.8\hsize]{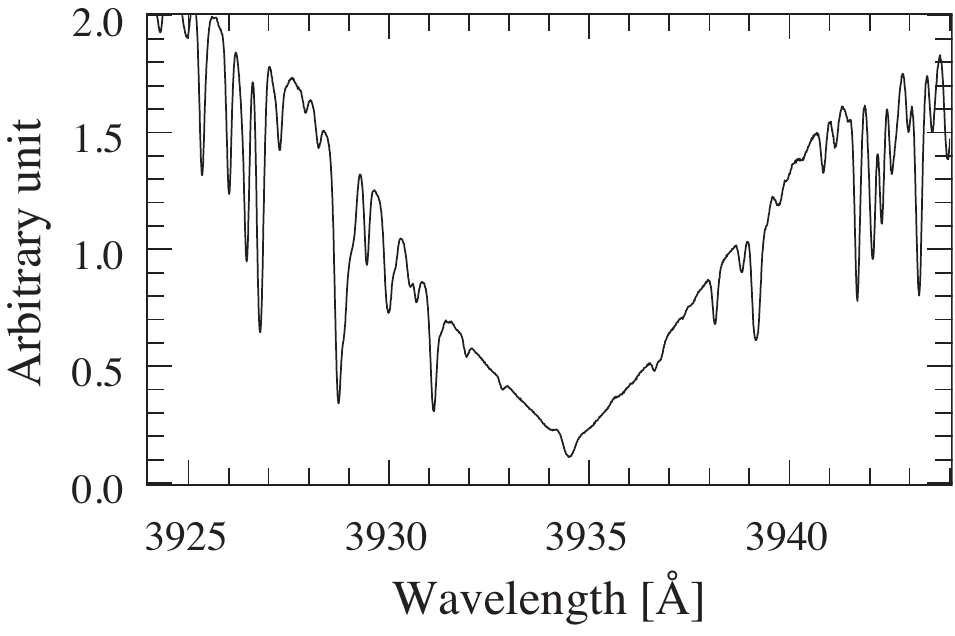}
    \caption{No emission in the \ion{Ca}{ii}~K line for the HD\,60532 spectra. This is the average spectrum of all the spectra used after recentering by the RV variations.}
    \label{fig:CaKline}
  \end{figure}

 In a (RV; bisector velocity span) diagram (Fig.~\ref{fig:bis_span}), the bisector velocity spans are spread mainly horizontally and are not correlated with the RVs. This argues in favor of a planetary origin to the variations rather than stellar activity (spots, pulsations). A spot origin can be rejected from several grounds: {\it 1)} given the star \vsini, much higher than the instrumental resolution (8\,\kms~compared to 3\,\kms), cool spots on the surface of the star would induce a correlation between bisector velocity spans and radial velocities (\cite{desort07} 2007), and the RV variations would have a period similar to the one of the star's rotation period\footnote{Here, the rotational period of the star $P_{\rm rot}$ is smaller than $\sim$7 days.}, whereas they are actually much longer; {\it 2)} also, spots able to produce such amplitude of RV variations would induce detectable photometric variations, whereas the photometry given by H{\small IPPARCOS} (\cite{hipparcos97} 1997) is constant with a scatter of only 0.004\,mag; {\it 3)} the star has a low level of activity (no emission in the \ion{Ca}{ii}~K line, Fig.~\ref{fig:CaKline}, and low $\log R'_{HK}$).\footnote{See online section for a figure showing the bisectors of all the spectra and for an example of an active, not young F6V star.}


  \subsection{Orbital parameters}
\begin{figure}[t!]
    \centering
    \includegraphics[width=1\hsize]{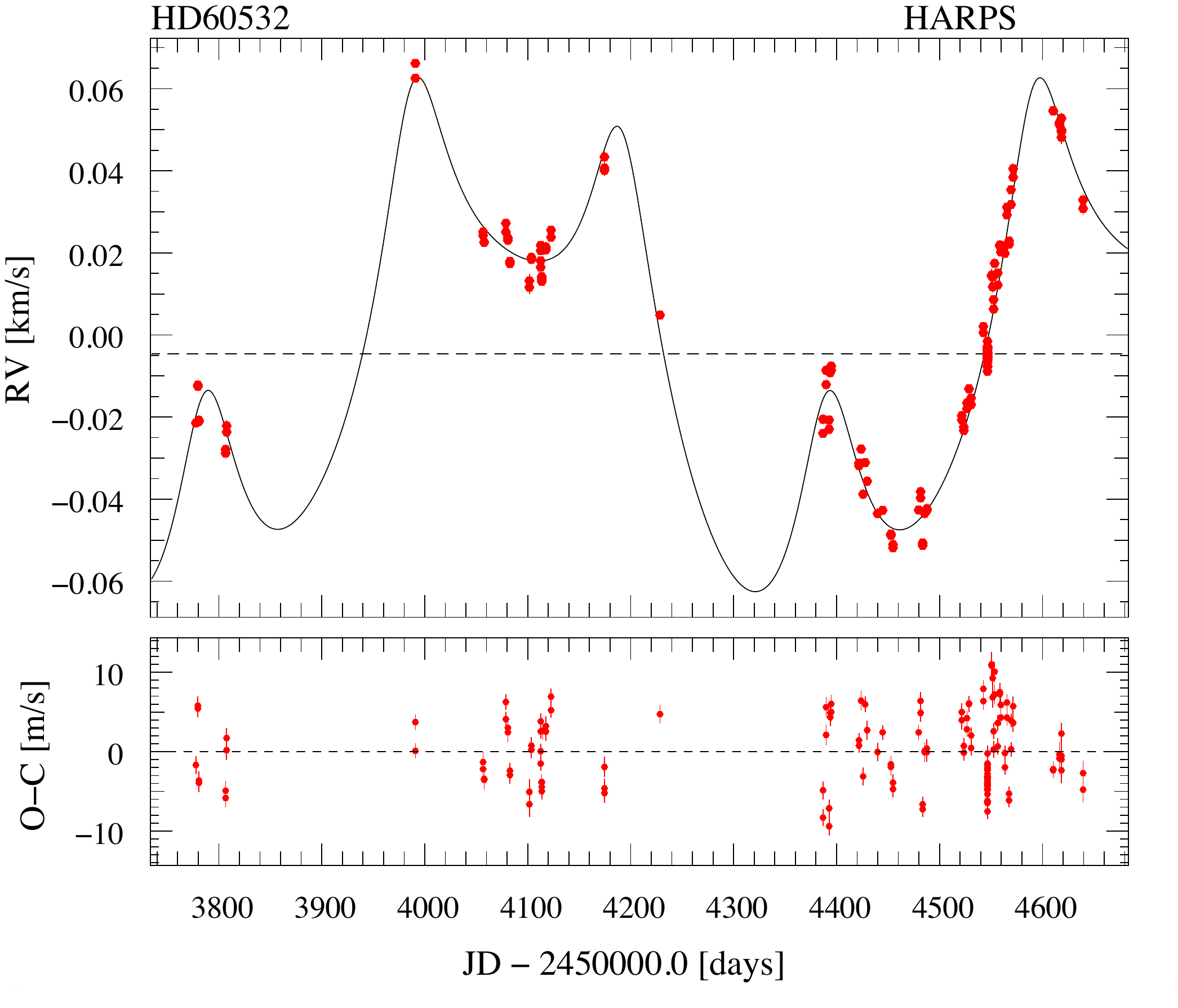}
    \includegraphics[width=0.9\hsize]{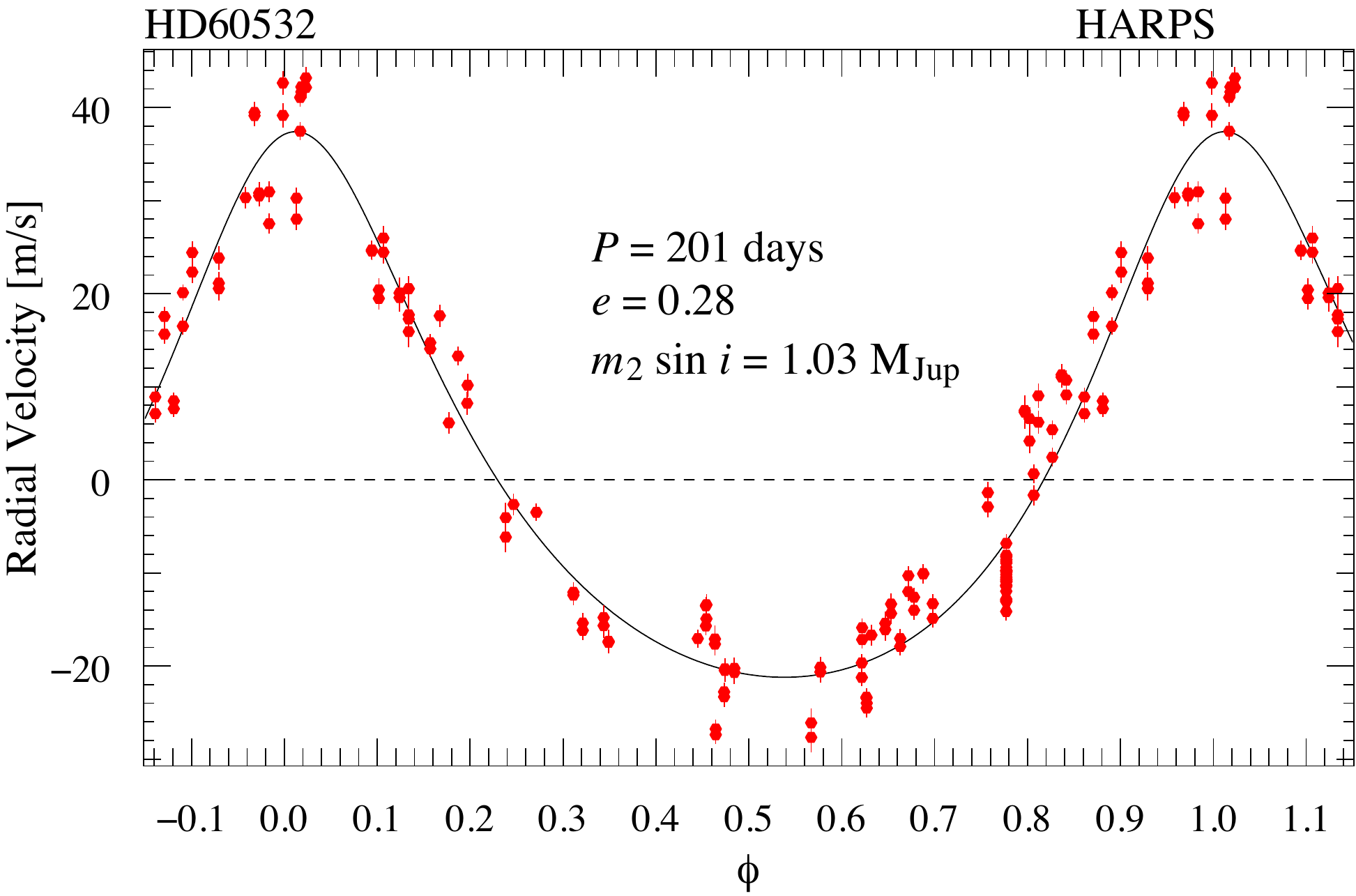}
    \includegraphics[width=0.9\hsize]{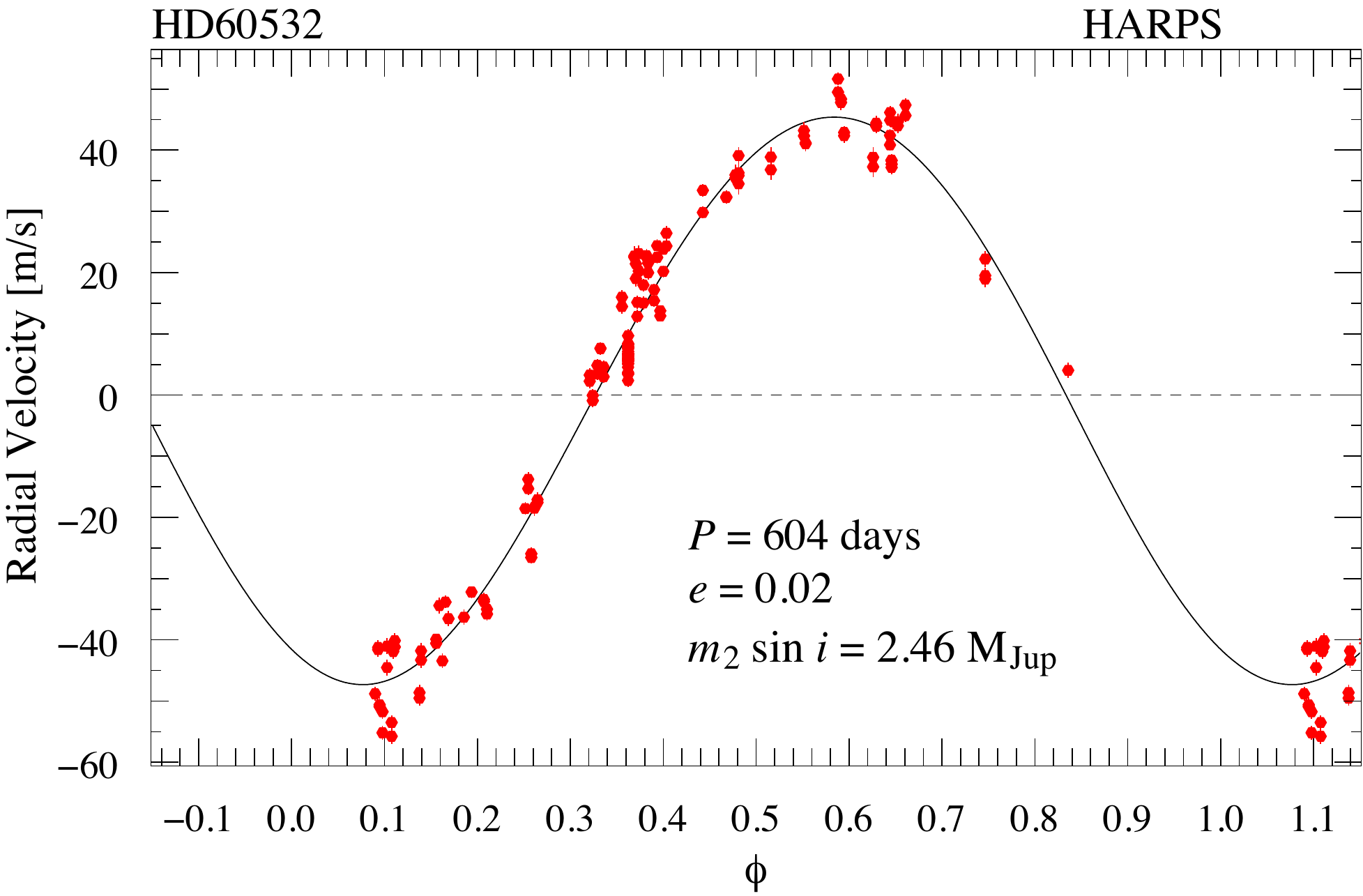}
    \caption{{\it Top}: \harps~radial velocities and orbital solution for HD\,60532, the panel below shows the residuals to the fitted orbital solution. {\it Middle and Bottom}: Phased fits on individual planets.}
    \label{fig:fit_2cc_free}
  \end{figure}

The observed RV can clearly not be fitted by one companion. The orbital parameters derived from the best two-companion Keplerian solution (Fig.~\ref{fig:fit_2cc_free}) are given in Table~\ref{tab:fit_2cc_free_param}. The residuals dispersion is 4\,\ms. A planetary system of two Jupiter-mass planets on long-period orbits fits the data very well. Assuming a primary mass of 1.44\,\Msun, this leads to a system with a 2.5-\Mjup~planet on a 604-day orbit and a 1.0-\Mjup~planet on a 201-day orbit.

 \begin{table}[t!]
    \caption{Best orbital solution for HD\,60532.}
    \label{tab:fit_2cc_free_param}
    \begin{center}
      \begin{tabular}{l l c c}
	\hline
        \hline
        Parameter        &                    & {\it Planet b}   & {\it Planet c} \\
        \hline
        $P$              & [days]             &  201.3 $\pm$ 0.6  &  604  $\pm$ 9  \\
        $T_0$            & [JD$-$2\,450\,000] &  3987  $\pm$ 2    &  3723 $\pm$ 158 \\
        $e$              &                    &  0.28  $\pm$ 0.03 &  0.02 $\pm$ 0.02\\
        $\omega$         & [deg]              &  $-$8.1 $\pm$ 4.9 &  $-$209 $\pm$ 92  \\
        $K$              & [\ms]              &  29.3  $\pm$ 1.4  &  46.4 $\pm$ 1.7 \\
        $N_{\rm meas}$   &                    &  147              & -- \\
        $\sigma_{O-C}$   & [\ms]              &  4.4              & -- \\
        reduced $\chi^2$ &                    &  4.4              & -- \\
        \hline
        $a_1\sin{i}$     & [$10^{-3}$\,AU]    &  0.52             &  2.6 \\
        $f(m)$           & [$10^{-9}$\,\Msun] &  0.46             &  6.2 \\
        $M_1$            & [\Msun]            & 1.44             & --  \\
        $m_2\sin{i}$     & [\Mjup]            &  1.03 $\pm$ 0.05  &  2.46 $\pm$ 0.09\\
        $a$              & [AU]               &  0.759 $\pm$ 0.001 &  1.58 $\pm$ 0.02 \\
        \hline
      \end{tabular}
    \end{center}
  \end{table}

  \subsection{Interpretation of the residuals}
  \label{sec:residuals}

\begin{figure}[t!]
  \centering
  \includegraphics[width=0.8\hsize]{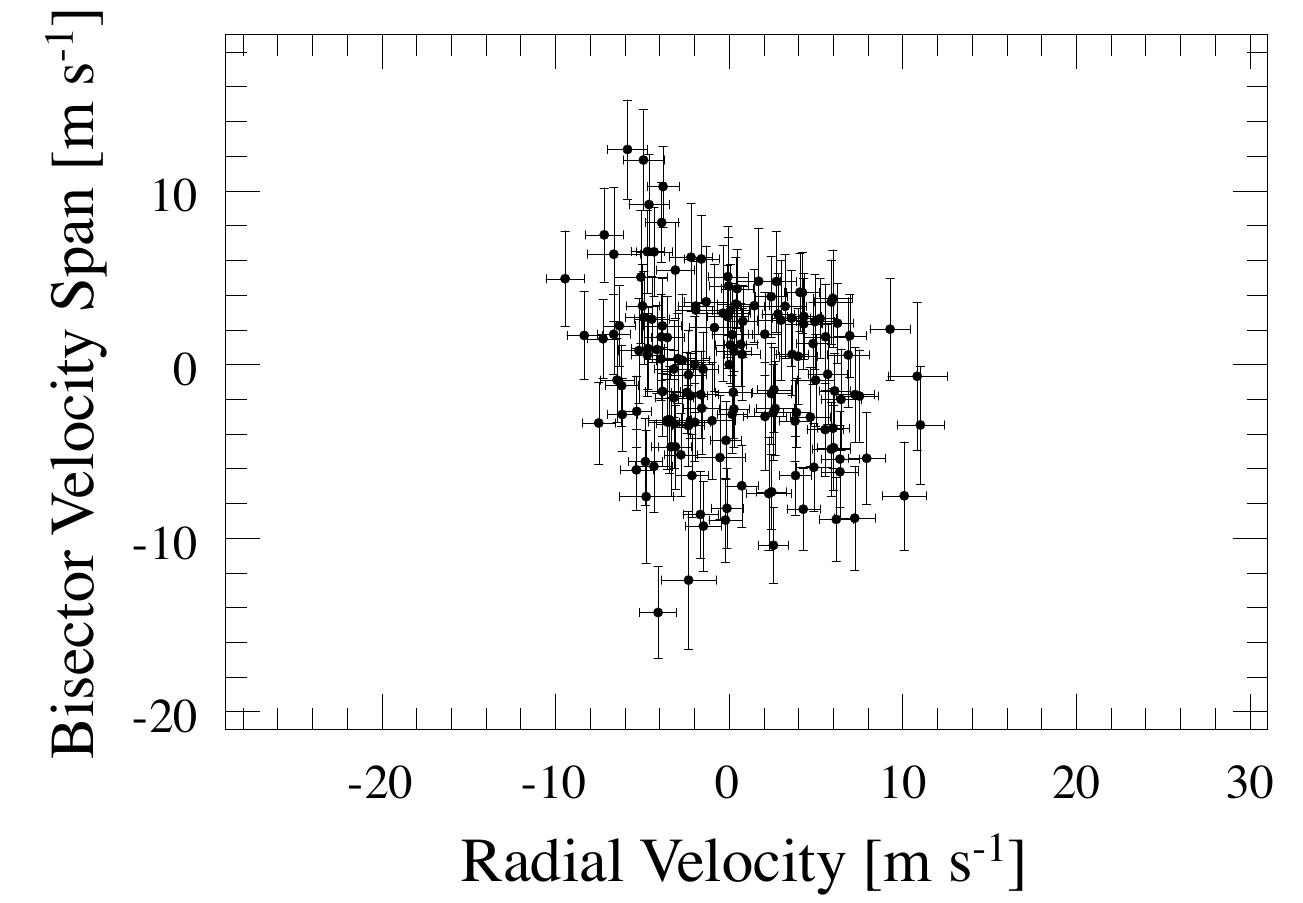}
  \caption{($O-C$; $O-C$ bisector velocity span) diagram: the $O-C$ bisector velocity spans are not horizontally spread (Sect.~\ref{sec:residuals}).}
  \label{fig:span_ominusc}
\end{figure}

  The residuals to the possible orbital solutions show variations, with standard deviation of 4.4\,\ms; we do not find any periodicity in these variations. The amplitude of variations is the same as the level of the low-amplitude variations seen in the line profile. Also, in a ($O-C$; $O-C$ bisector velocity span) diagram, the $O-C$ bisector velocity spans are not horizontally spread, which shows that these remaining variations are not due to the presence of a smaller-mass planet (Fig.~\ref{fig:span_ominusc}). They are then probably due to stellar intrinsic phenomena. We monitored the star continuously for one hour (28 spectra) and measured RV variations with a peak-to-peak amplitude of 9\,\ms. This corresponds to the observed semi-amplitude of the residuals. Such high-frequency variations argue for stellar pulsations and if we would like to increase the measurement accuracy we would need to average this effect. Finally, we could not find any periodicity associated with the residuals.

\section{Dynamical issues}

The two orbital periods derived from the fit suggest a possible 3:1 mean-motion resonance between the two planets. Here we investigate the dynamical behaviour of the system to possibly distinguish between resonant and non-resonant configurations. As we show below, the uncertainty on the semi-major axis values for planets {\it b} and {\it c} (mainly on Planet~{\it c}, see Table~\ref{tab:fit_2cc_free_param}) does not permit to state whether the planets are actually locked in mean-motion resonance of just next to it.

The two planets are said to be in mean-motion resonance when their orbital periods achieve a simple rational ratio. More generally, they will be assumed to be locked in a $p+q:p$, where $p$ and $q$ are integers. $|q|$ is called the \emph{order} of the resonance, and denotes the number of stable conjunction positions. In the 3:1 case, we have $q=2$. The dynamics within a resonance is usally described via the use of the following variable called \emph{critical argument of the resonance}:

\begin{equation}
\sigma = \frac{p+q}{q}\,\lambda_c-\frac{p}{q}\,\lambda_b-\varpi_b\qquad,
\label{sg00}
\end{equation}
where $\lambda_b$ and $\lambda_c$ are the mean longitudes of the inner and the outer planet, respectively, and where $\varpi_b$ is the longitude of periastron of the inner planet (here the less massive one). Resonant orbits are characterized by a libration of $\sigma$ around an equilibrium position $\sigma_0$, while non-resonant orbit exhibit a circulation of $\sigma$. In non-resonant configurations, the semi-major axes are nearly secular invariants, due to independent phase averaging over both orbits, while in resonant configurations, the $\sigma$ libration induces secular eccentricity and semi-major axis oscillations.

We integrate the 3-body system (star + 2 planets, assuming $\sin{i}=1$) starting from the today fitted orbital solution, using the symplectic N-body code SyMBA (\cite{duncan98} 1998). The use of a powerful symplectic code allows to adopt a large timestep while remaining accurate. We adopt here a timestep of 0.025\,yr, \ie a bit less that 1/20 of the smallest orbital period. This is a standard prescription in the use of this code to ensure a relevant accuracy. We also made tests using a ten times smaller timestep to check the validity of our integrations. From a dynamical point of view the orbital solution as given in Table~\ref{tab:fit_2cc_free_param} is not complete. The longitude of nodes $\Omega_b$ and $\Omega_c$ are not constrained. Note that thanks to rotational invariance, only the difference $\Omega_c-\Omega_b$ is a relevant parameter. It will be treated as a free parameter in our integrations.

We first focus on resonant configurations. For this we assume for the semi-major axes $a_b=0.759$\,AU and $a_c=1.57928$\,AU. This choice for $a_c$ (within the error bar) ensures an exact resonant configuration.

\begin{figure*}
\makebox[\textwidth]{
\includegraphics[width=0.5\hsize,origin=br]{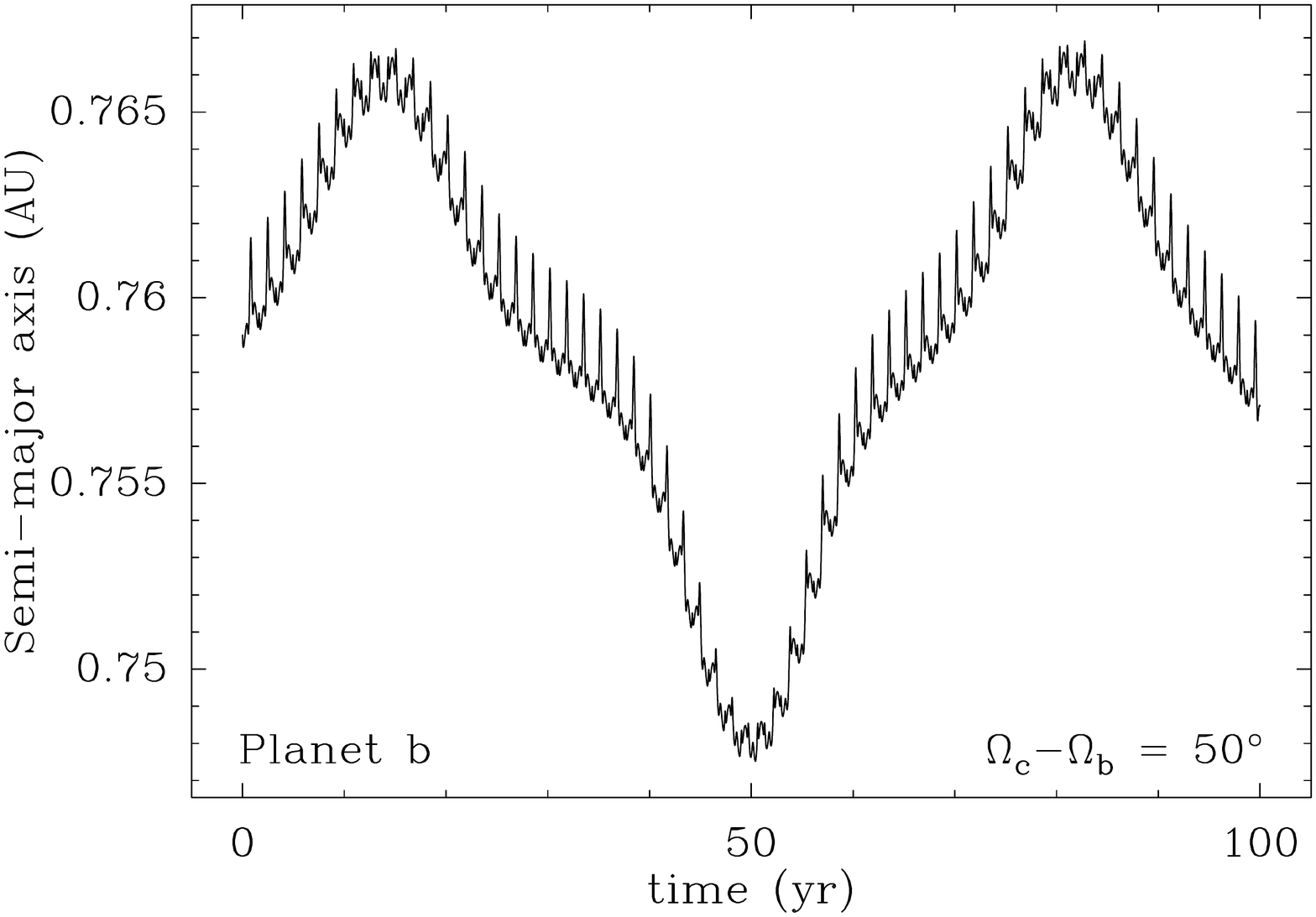}\hfil
\includegraphics[width=0.5\hsize,origin=br]{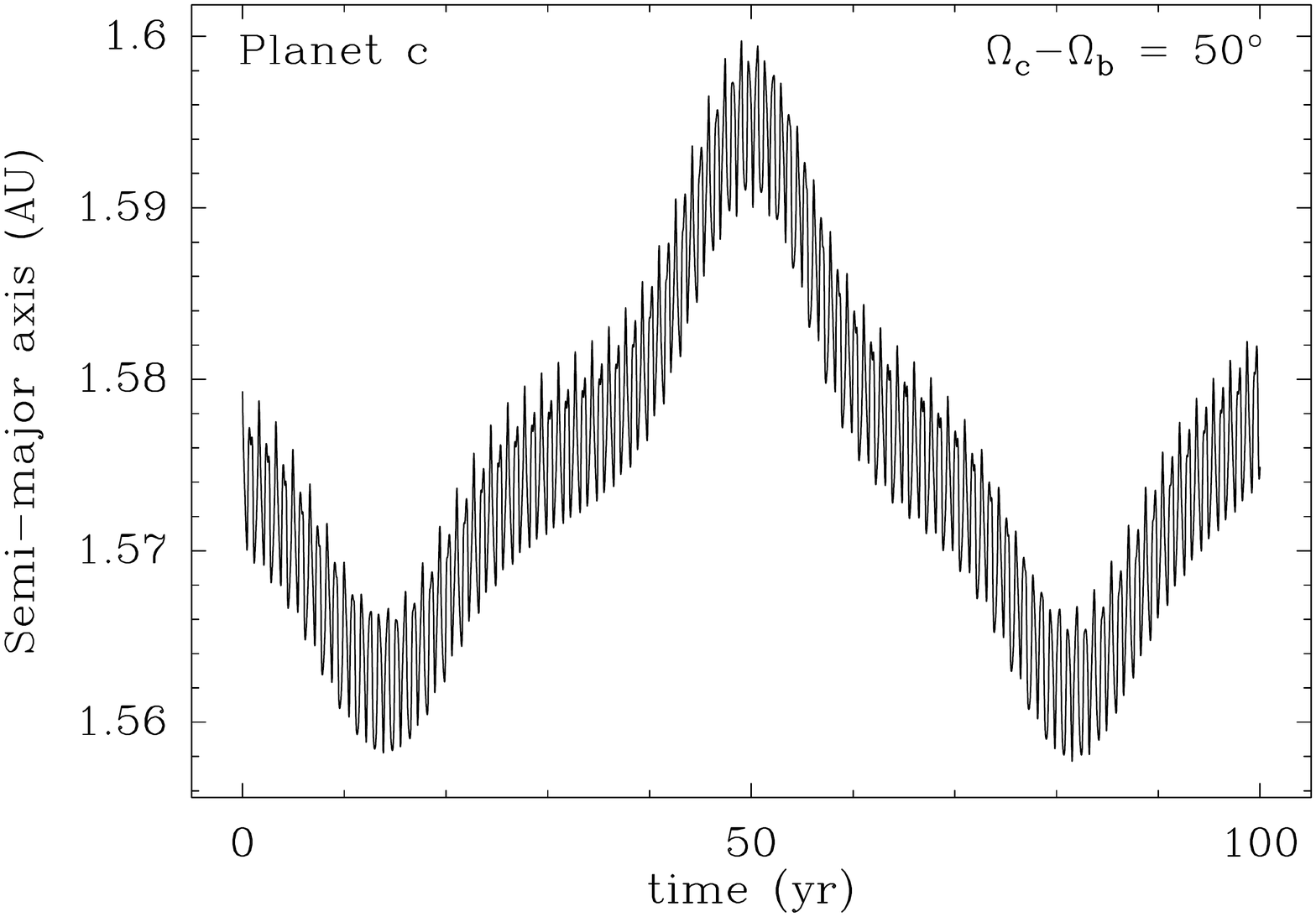}}
\makebox[\textwidth]{
\includegraphics[width=0.5\hsize,origin=br]{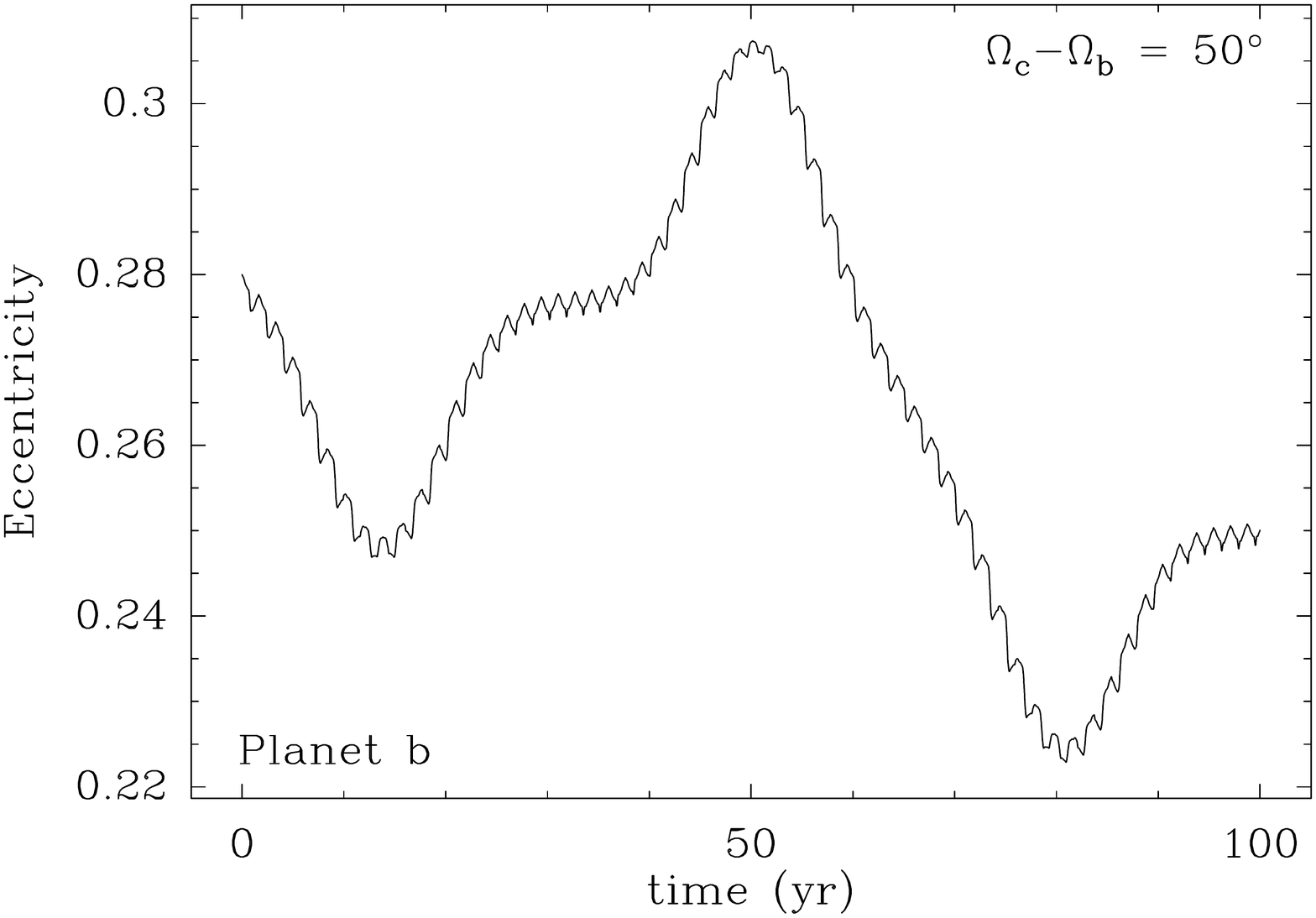}\hfil
\includegraphics[width=0.5\hsize,origin=br]{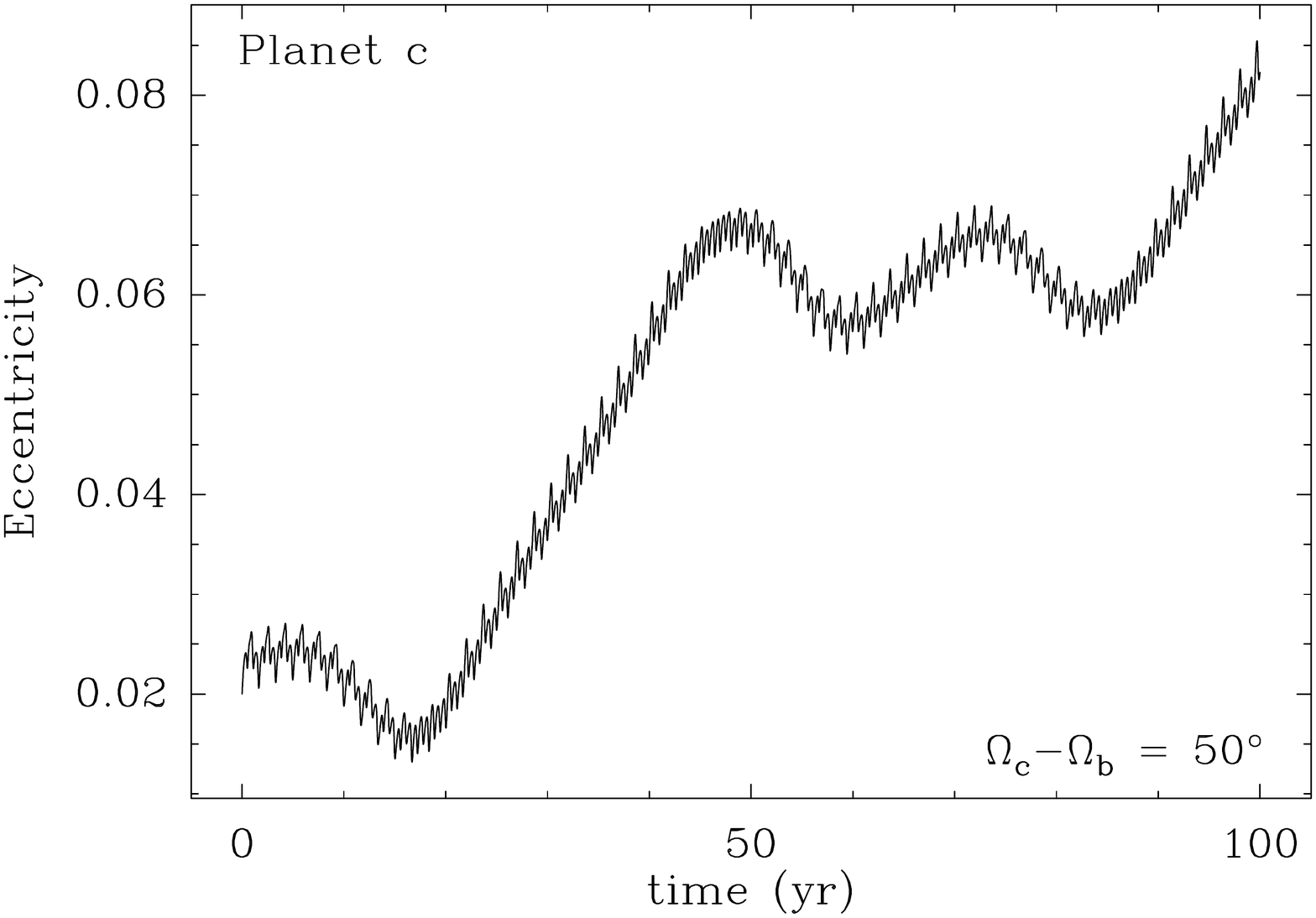}}
\caption{Orbital evolution over 100\,yr of the semi-major axes (top) and eccentricities (bottom) for planets {\it b} (left) and {\it c} (right), under their mutual perturbations, in a 3:1 resonance configuration, for an initial choice of $\Omega_c-\Omega_b=50\degr$.}
\label{res_100yr}
\end{figure*} 

\begin{figure}
\includegraphics[width=\hsize]{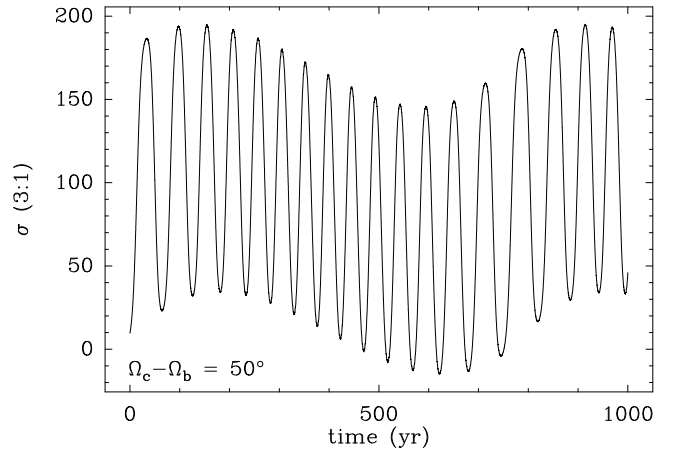}
\caption{Evolution of the 3:1 critical argument $\sigma$ over 1000\,yr, in the same condition as described in Figure~\ref{res_100yr}. We note the $\sigma$-libration characteristic for resonant motion.}
\label{sigma_1000yr}
\end{figure}

Figure~\ref{res_100yr} shows the orbital evolution over 100\,yr of the two planets in this configuration under their mutual perturbations, for a choice of $\Omega_c-\Omega_b=50\degr$. We note that both the semi-major axes and the eccentricities exhibit a superimposition of two types of variations: we first have a long period oscillation with $\sim 70$\,yr period, in addition to high frequency (1--2\,yr periodicity), but much smaller amplitude changes. The former oscillation is related to the resonant libration motion (see Fig.~\ref{sigma_1000yr}), while the latter is phased with the synodic motion of the two planets. It is thus related to the mutual perturbations of the two bodies at conjunction. The fact that this high-frequency term shows up here is an indication for weak -- but significant -- chaos in the system. The integration was extended up to $10^8$\,yr. Over this time span, the planetary system does not show any indication for instability. The semi-major axes of the two planets oscillate between 0.748\,AU and 0.767\,AU for Planet {\it b}, and between 1.558\,AU and 1.6\,AU for Planet {\it c}. The eccentricity of Planet {\it b} oscillates between 0.13 and 0.32, and that of Planet {\it c} between 0 and 0.14. Note that the present value of Planet {\it c}'s eccentricity is close to the bottom of its actual variation range.

\begin{figure*}
\makebox[\textwidth]{
\includegraphics[width=0.5\hsize]{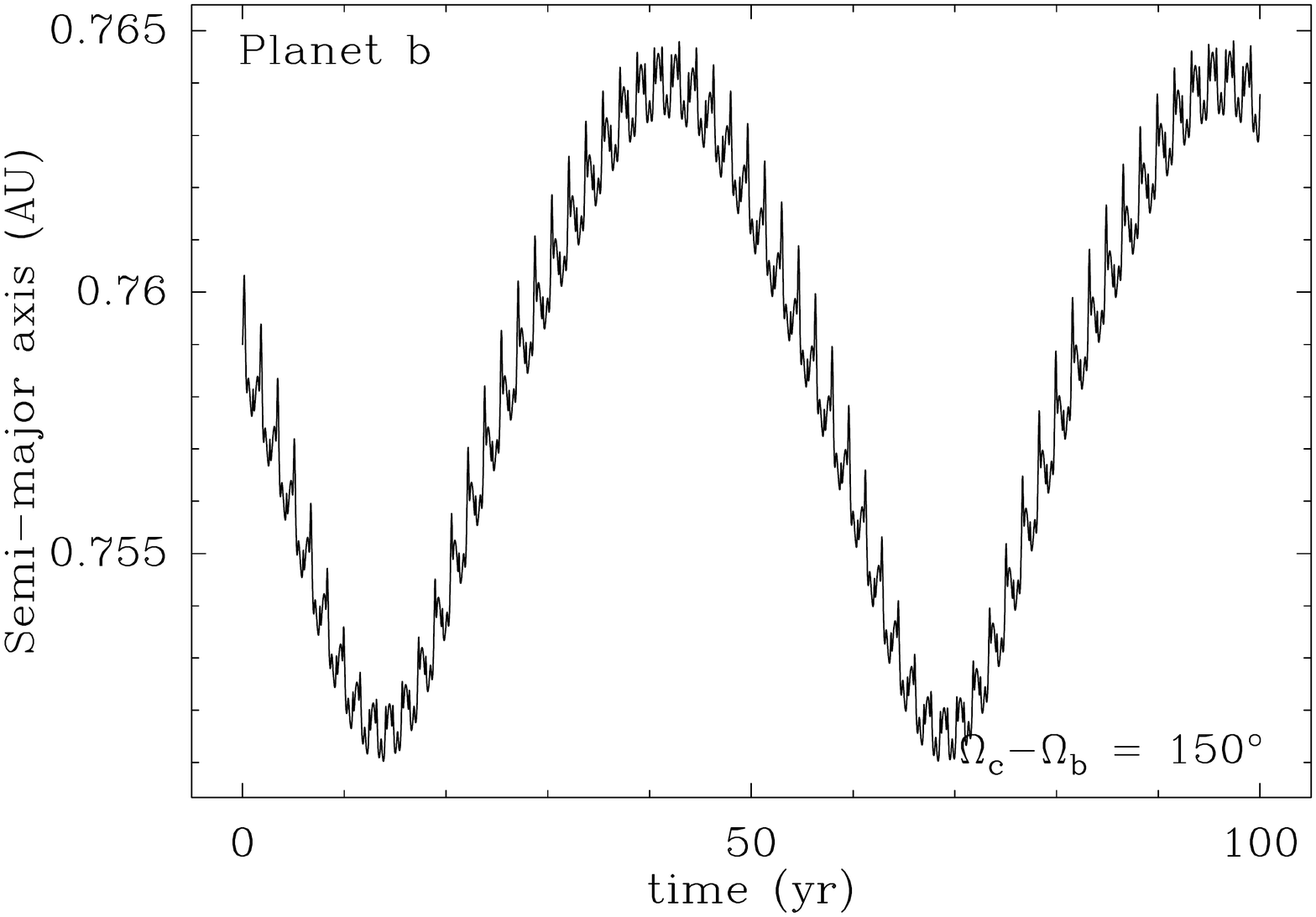}\hfil
\includegraphics[width=0.5\hsize]{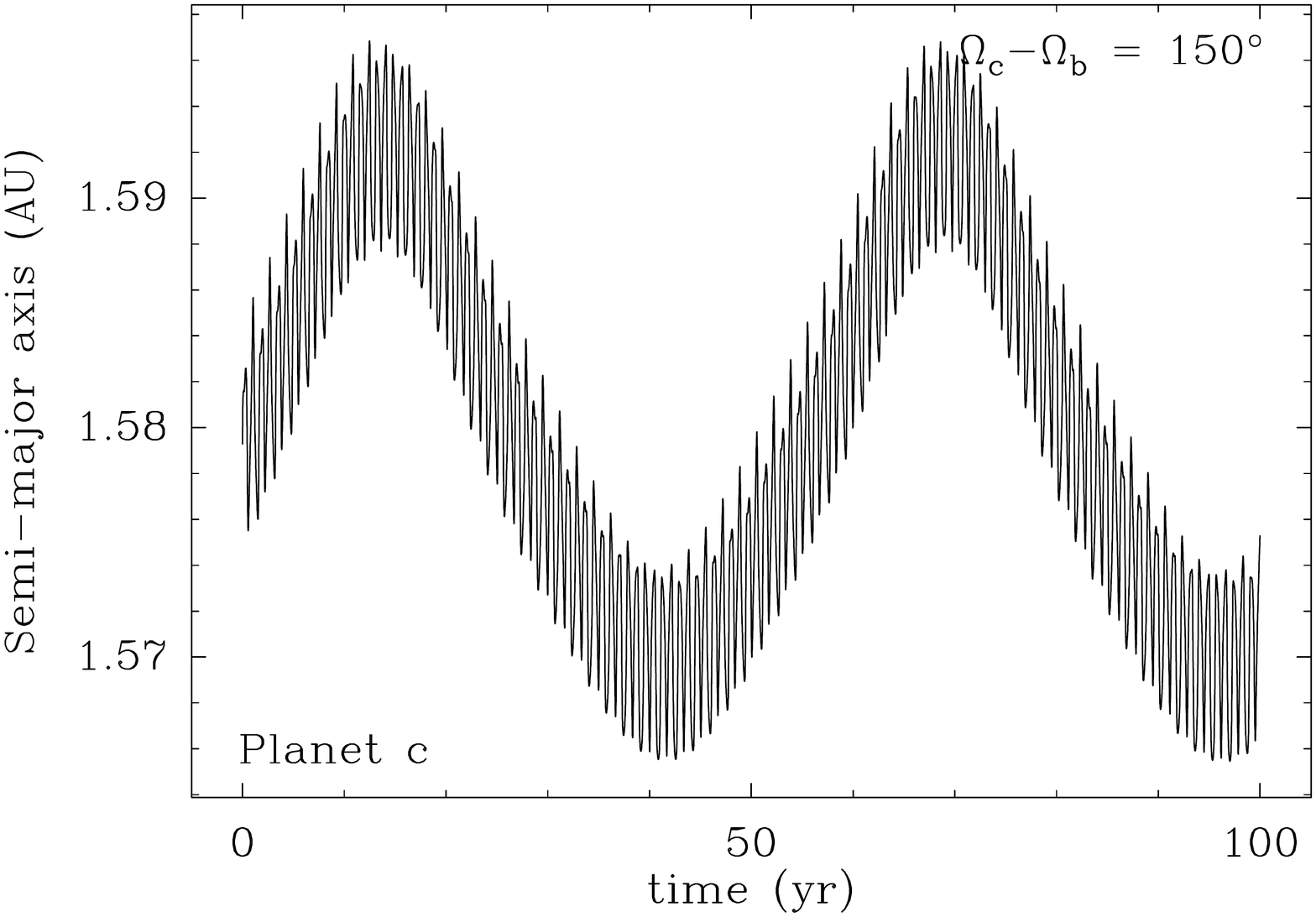}}
\caption{Evolution of the semi-major axes of the two planets in the same condition as described in Fig.~\ref{res_100yr}, except that we assumed $\Omega_c-\Omega_b=150\degr$.}
\label{res_150}
\end{figure*}

Figure~\ref{res_100yr} shows that, given the error bars listed in Table~\ref{tab:fit_2cc_free_param}, the secular evolution of the semi-major axis of Planet {\it b} should be detectable within $\sim 10$\,yr from now. In fact, what is expected to be detected is only the low frequency secular variation. The high frequency oscillation remains below the error bar.

Now, we point out that Fig.~\ref{res_100yr} corresponds to $\Omega_c-\Omega_b=50\degr$. Figure~\ref{res_150} shows the evolution of the semi-major axes in the same conditions, but starting with $\Omega_c-\Omega_b=150\degr$. We still see the same kind of oscillations, but phased differently. In Fig.~\ref{res_100yr}, the semi-major axis of Planet {\it b} increases in the first 10 years, while in Fig.~\ref{res_150}, it decreases.

\begin{figure*}
\makebox[\textwidth]{
\includegraphics[width=0.5\hsize]{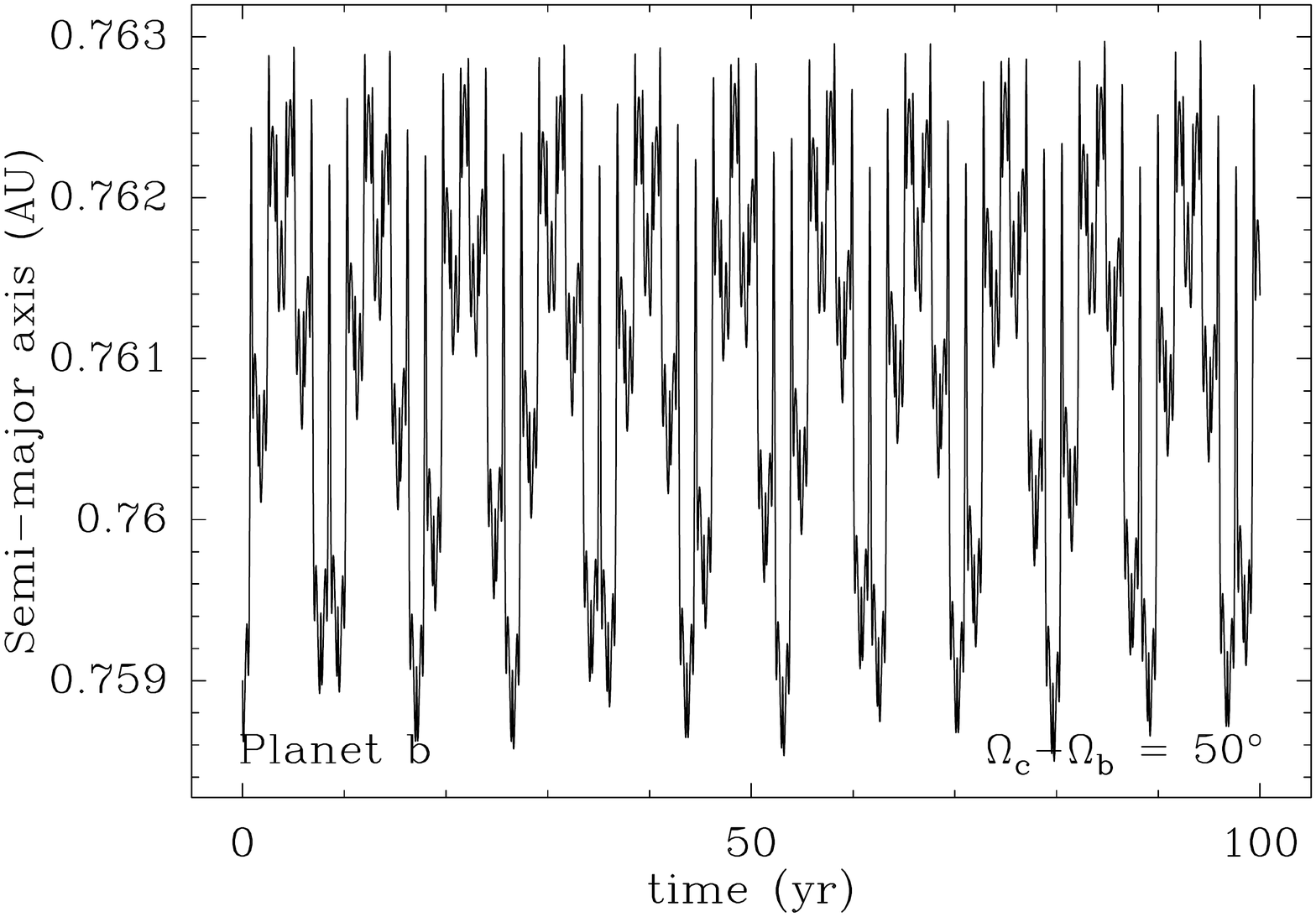}\hfil
\includegraphics[width=0.5\hsize]{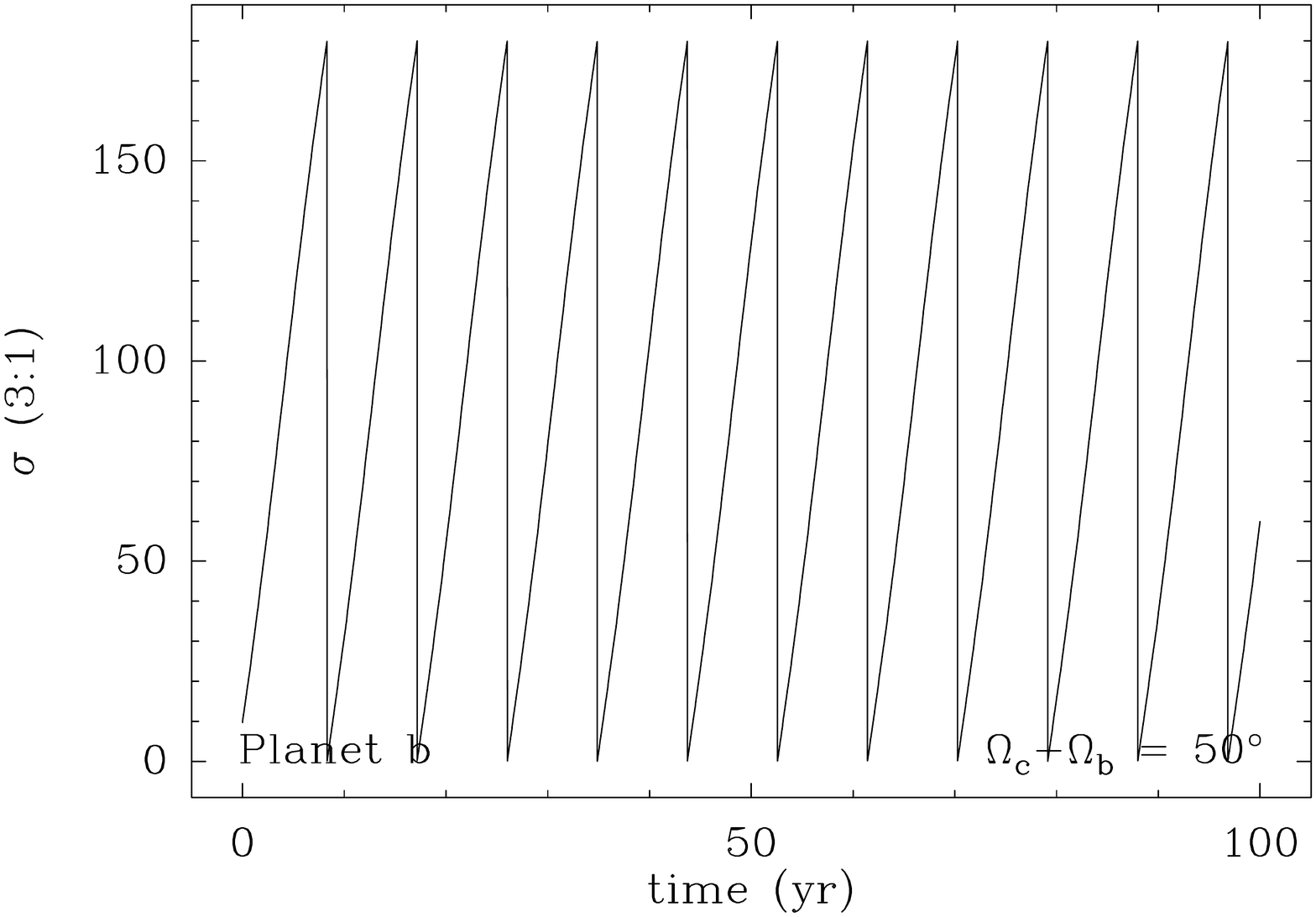}}
\caption{Evolution of the semi-major axis of Planet {\it b} and of the critical angle $\sigma$ for 3:1 resonance in the same conditions as described in Fig.~\ref{res_100yr}, but with $a_c=1.57528\,$AU. This is a non-resonant configuration.}
\label{nonres}
\end{figure*}

Although the actual value of $\Omega_c-\Omega_b$ is unknown, the detection of semi-major axis evolution of Planet {\it b} within 10 years should be considered as a strong indication of a resonant configuration. Figure~\ref{nonres} shows the evolution of the semi-major axis of Planet {\it b} and of the critical angle $\sigma$ for 3:1 resonance for an initial choice of $a_c=1.57528$\,AU. As can be seen from the evolution (circulation) of $\sigma$, this configuration is non-resonant, while still within the error bar of the orbital fit. Hence from the orbital fit itself, it is not possible to definitely state whether the two planets are actually locked in 3:1 resonance or not. From Fig.~\ref{nonres}, we see that in non-resonant configuration, the variations of the semi-major axis of Planet {\it b} achieve a much lower amplitude than in the resonant case. This remains true for any choice of initial non-resonant configuration. In fact, in that case, we only have the high-frequency, small-amplitude oscillation. As expected in non-resonant case, there is no secular evolution of the semi-major axis, contrary to the resonant case. The net consequence of this is that given the error bars of the fit, no semi-major axis variation for Planet {\it b} should be detected within 10 years in the non-resonant case.

Our conclusions concerning this dynamical study are thus the following:
\begin{itemize}
\item Given the error bars of the fit, it is not possible to definitely state the planets are actually locked in mean-motion resonant or not;
\item The two-planet system is dynamically stable although significantly chaotic;
\item Although it is impossible to predict the variation sense due to unconstrained $\Omega_c-\Omega_b$, any detection of variation in the semi-major axis of Planet {\it b} within 10 years from now should be a strong indication for a resonant configuration. Further monitoring of this system should therefore be initiated to detect this possible variation.
\end{itemize}

\section{Concluding remarks}

  We have shown that HD\,60532, an F6IV--V, 1.44\,\Msun~star hosts two planets with minimum masses of 1 and 2.5\,\Mjup~and orbital separations of 0.76 and 1.58\,AU respectively, in a possible 3:1 resonance which needs to be confirmed within the next 10 years. Noticeably sofar, only one other multiple system had been reported around a MS star more massive than 1.3\,\Msun~(\object{HD\,169830}; 1.4\,\Msun). Note also that the low metallicity of HD\,60532 is not common for stars harbouring Jupiter-mass planets; the relation between the star's metallicity and the presence of massive planets has still to be investigated further.

\begin{acknowledgements}
  We acknowledge financial support from the French Agence Nationale pour la Recherche, ANR grant NT05-4\_44463. We are grateful to ESO and the La Silla Observatory for the time allocation and to their technical staff. We also thank the HARPS GTO observers for performing some observations, G\'erard Zins and Sylvain C\`etre for their help in implementing the SAFIR interface. These results have made use of the SIMBAD database, operated at CDS, Strasbourg, France.
\end{acknowledgements}

\begin{appendix}
\section{Additional material}

\begin{figure}[h!]
    \centering
    \includegraphics[width=0.8\hsize]{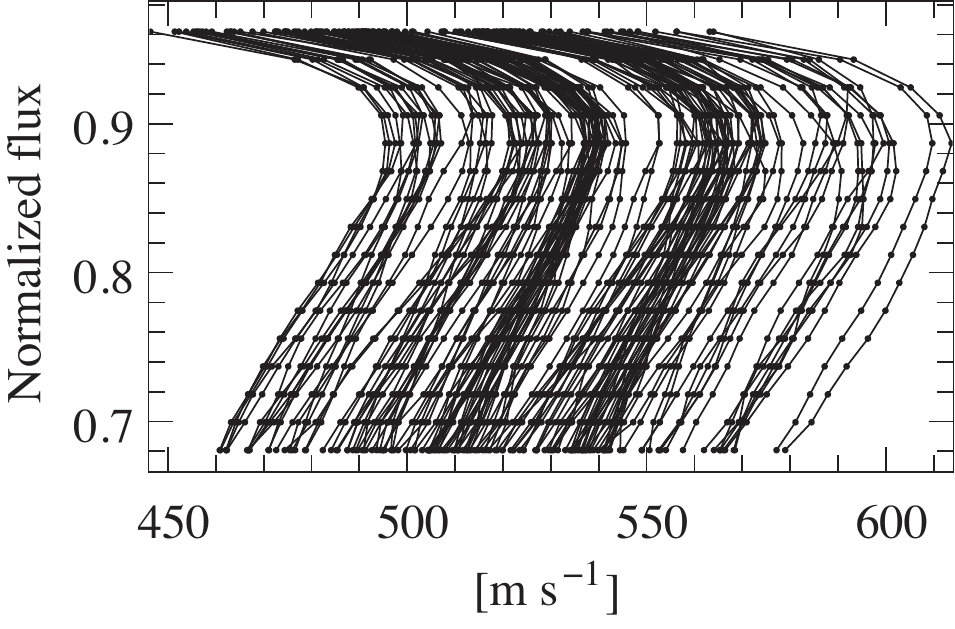}
    \caption{Bisectors of the CCFs.}
    \label{fig:bisectors}
  \end{figure}

Figure~\ref{fig:bisectors} presents the bisectors of the CCFs (which stands for all the lines) for the spectra from which we measured the radial velocities of HD\,60532. It shows that spectra are mainly shifted, with yet small line-profile variations superimposed to the shifts, and thus that the radial velocities measured are not induced by line-profile variations.

\begin{figure}[h!]
    \centering
    \includegraphics[width=0.8\hsize]{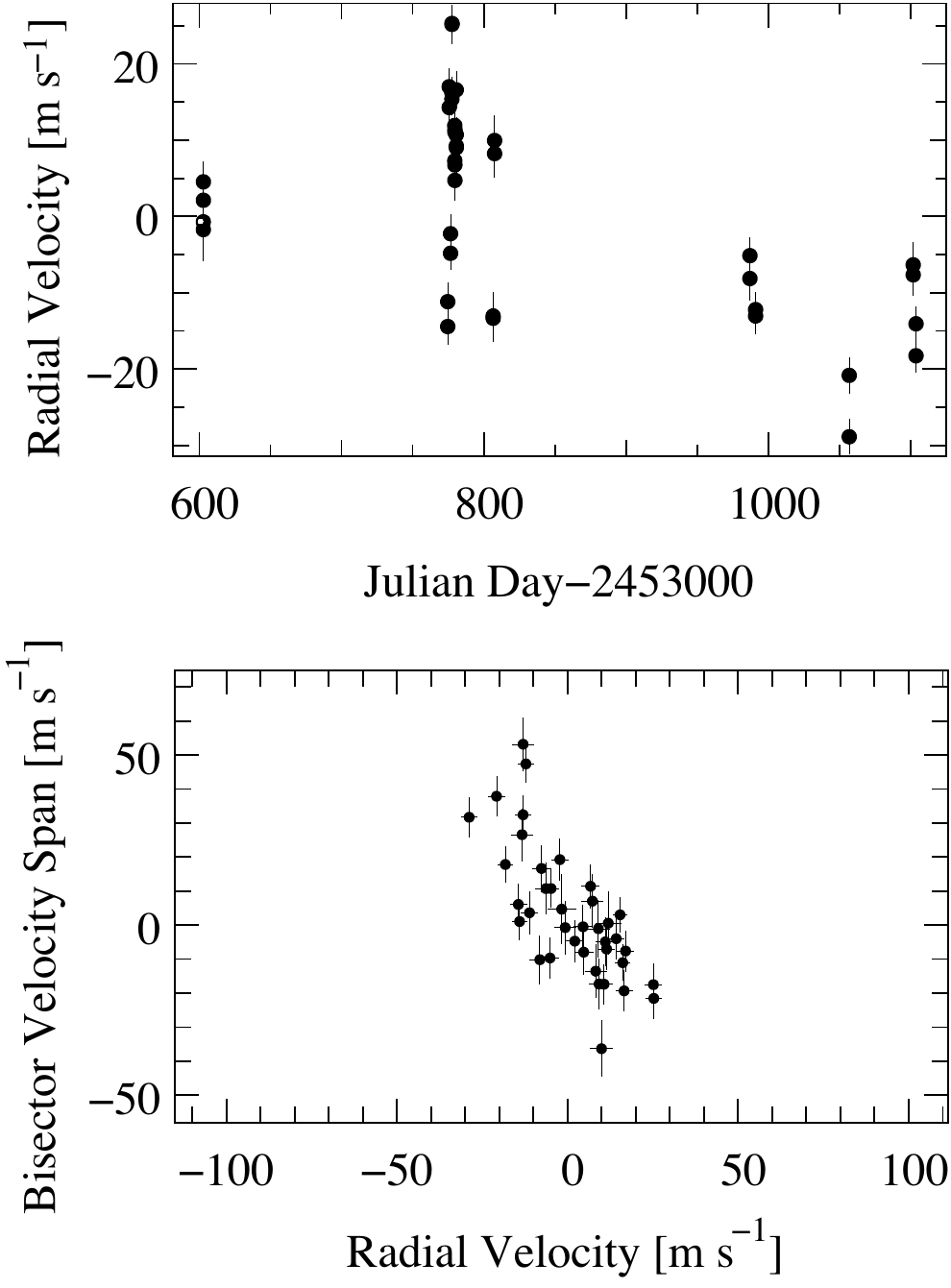}
    \caption{Example of an active F6V star for which measured RVs ({\it top}) which are correlated to the bisector velocity spans ({\it bottom}).}
    \label{fig:hd30}
  \end{figure}

Figure~\ref{fig:hd30} shows an example of an active F6V star (\object{HD\,30652}, \vsini\,=\,16\,\kms) that we observed during the same survey, and for which RV measurements are correlated with line-profile variations, indicating that they are are induced by a spot.

\end{appendix}


\begin{thebibliography}{}
\bibitem[Chelli]               {chelli00}    Chelli, A., 2000, A\&A 358, L59
\bibitem[Desort et al.]        {desort07}    Desort, M., Lagrange, A.-M., Galland, F., et al., 2007, A\&A 473, 983
\bibitem[Duncan et al.]        {duncan98}    Duncan, M.~J., Levison, H.~F. and Lee, M.~H., 1998, AJ, 116, 2067
\bibitem[Galland et al.]       {galland05a}  Galland, F., Lagrange, A.-M., Udry, S., et al., 2005a, A\&A 443, 337
\bibitem[Gray et al.]          {gray06}      Gray, R.~O., Corbally, C.~J., Garrison, R.~F., et al., 2006, AJ, 132, 161
\bibitem[ESA]                  {hipparcos97} ESA 1997, The Hipparcos and Tycho Cat, ESA SP-1200
\bibitem[Hoffleit et al.]      {hoffleit91}  Hoffleit, D., Warren Jr, W.H., 1991, Bright Star Catalogue (5th Revised Ed.), NSSDC/ADC
\bibitem[Lagrange et al.]      {lagrange08}  Lagrange, A.-M., Desort, M., Galland, F., et al., 2008, A\&A, submitted
\bibitem[Lovis \& Mayor]       {lovis07}     Lovis, C., Mayor, M., 2007, A\&A 472, 657
\bibitem[Mayor et al.]         {mayor03}     Mayor, M. et al, 2003, The Messenger 114, 20
\bibitem[Nordstr\"{o}m et al.] {nordstrom04} Nordstr\"{o}m B., Mayor, M., Andersen J., et al., 2004, A\&A 418, 989
\bibitem[Sato et al.]          {sato05}      Sato, B., Kambe, E., Takeda, Y., et al. 2005, PASJ 57, 97
\end{thebibliography}
\end{document}